\documentclass[useAMS]{mn2e}
\usepackage{graphicx}
\usepackage{txfonts}

\title[Stellar sources of dust in the high redshift Universe]{Stellar sources of dust in the high redshift Universe}
\author[R. Valiante, R. Schneider, S. Bianchi and A. C. Andersen]{Rosa Valiante$^{1}$\thanks{E-mail:
valiante@arcetri.astro.it}, Raffaella Schneider$^{2}$, Simone Bianchi$^{2}$ and Anja C. Andersen$^{3}$ \\
$^{1}$Dipartimento di Astronomia, Universita' di Firenze, Largo Enrico Fermi 2, 50125 Firenze, Italy\\
$^{2}$INAF - Osservatorio Astrofisico di Arcetri, Largo Enrico Fermi 5, 50125, Firenze, Italy\\
$^{3}$Dark Cosmology Centre, Niels Bohr Institute, University of Copenhagen, Juliane Maries Vej 30, DK-2100 Copenhagen, Denmark}
\begin{document}

\date{Accepted . Received }

\pagerange{\pageref{firstpage}--\pageref{lastpage}} 
\pubyear{2009}

\maketitle

\label{firstpage}

\begin{abstract}
With the aim of investigating whether stellar sources can account for the $\geq 10^8 M_{\odot}$ dust masses inferred from mm/sub-mm observations
of samples of $5<z<6.4$ quasars, we develop a chemical evolution model which follows the evolution of metals and dust on the stellar 
characteristic lifetimes, taking into account dust destruction mechanisms. 
Using a grid of stellar dust yields as a function of the initial mass and metallicity over the range $1 - 40 M_\odot$ 
and $0 - 1 Z_\odot$, we show that the role of AGB stars in cosmic dust evolution at high redshift might have been over-looked. 
In particular, we find that (i) for a stellar population forming according to a present-day Larson initial mass function (IMF) with 
$m_{ch} = 0.35 M_{\odot}$, the characteristic timescale at which AGB stars dominate dust production ranges between 150 and  500~Myr, 
depending both on the assumed star formation history and on the initial stellar metallicity; (ii) this result is only moderately 
dependent on the adopted stellar lifetimes, but it is significantly affected by variations of the IMF: for a $m_{ch} = 5 M_\odot$, 
dust from AGB starts to dominate only on timescales larger than 1 Gyr and SNe are found to dominate dust evolution when $m_{ch} \geq 10 M_\odot$. 
We apply the chemical evolution model with dust to the host galaxy of the most distant quasar at $z = 6.4$, SDSS J1148+5251. 
Given the current uncertainties on the star formation history of the host galaxy, 
we have considered two models: (i) the star formation history obtained in a numerical simulation by Li et al. (2007) 
which predicts that a large stellar bulge is already formed at $z = 6.4$, 
and (ii) a constant star formation rate of $1000 M_\odot$/yr, as suggested by the observations if most of the FIR luminosity is due to young stars. 
The total mass of dust predicted at $z = 6.4$ by the first model is $2 \times 10^8 M_\odot$, within the range of values inferred by observations, 
with a substantial contribution ($\sim 80$\%) of AGB-dust. When a constant star formation rate is adopted, the contribution of AGB-dust decreases to $\sim 50$\% but the
total mass of dust formed is a factor 2 smaller. Both models predict a rapid enrichment of the ISM with metals and a relatively mild evolution of the carbon abundance,
in agreement with observational constraints. This supports the idea that stellar sources can account for the dust observed but show that
the contribution of AGB stars to dust production cannot be neglected, even at the most extreme redshifts currently accessible to observations. 
\end{abstract}


\section{Introduction}

It is estimated that 30\% or more of the light emitted from stars
in the Universe is absorbed and re-emitted by dust in the infrared 
(IR). The cosmic IR background, therefore, records the cumulative 
IR emission from galaxies at all redshifts and provides an important 
constraint on the global history of star formation. This emission is 
affected by dust evolution in galaxies, in particular at high redshift 
($z > 5$), where the presence of dust can influence the galaxy number 
counts and the evolution of quasars (e.g. obscuring star-forming 
galaxies or Active Galactic Nuclei;  Smail, Ivison \& Blain 
1997; Hughes et al. 1998, Maiolino et al. 2003a). 

Evidence for the presence of dust at high redshifts comes from 
observations of damped Ly$\alpha$  systems (Pettini et al. 1994; Prochaska 
\& Wolfe 2002; Ledoux, Bergeron \& Petitjean 2002) and from the 
detection of dust thermal emission from high redshift quasars (QSOs) selected 
from the Sloan Digital Sky Survey (SDSS) re-observed at millimetre wavelengths (Omont et al. 2001; 
Carilli et al. 2001; Bertoldi \& Cox 2002). The inferred far-IR (FIR) 
luminosities of samples of $5 < z < 6.4$ quasars are consistent with 
thermal emission from warm dust ($T < 100$~K), with dust masses in 
excess of $10^8 M_{\odot}$  (Bertoldi et al. 2003; Robson et al. 2004; Beelen 
et al. 2006). Despite the fact that these high redshift 
quasars are rare objects, hardly representative of the dominant 
star forming galaxy population, the FIR and sub-mm observations 
indicate that early star formation leads to rapid enrichment of their 
host galaxies Interstellar Medium (ISM) with metals and dust. 
This is consistent with the super-solar metallicities inferred from the optical 
emission-line ratios for many of these systems (Pentericci et 
al. 2002; Freudling et al. 2003; Maiolino et al. 2003b). 

In present-day galaxies, the major sources of interstellar dust 
are believed to be low and intermediate-mass evolved stars during 
the Asymptotic Giant Branch (AGB) phase. Since these stars take 
a long time to evolve to their dust producing stages, it is commonly 
believed that they cannot be responsible for all of the dust seen 
at high redshifts. In fact, AGB stellar lifetimes ($10^8$ to $10^9$~yr) 
are comparable to the age of Universe at redshift $\geq 6$ (Morgan \& 
Edmunds 2003; Marchenko 2006). As a result, if the observed dust is 
produced by stellar processes, supernova (SN) ejecta appear to be 
the only viable sites of grain condensation fast enough to explain 
these large dust masses.

This scenario has been tested through observations of the reddened quasar 
SDSSJ1048+46 at $z = 6.2$ (Maiolino et al. 2004) and of the Spectral Energy 
Distribution (SED) of the GRB 050904 afterglow (Stratta et al. 2007). 
In both sources, the inferred dust extinction curve is different with respect 
to any of the extinction curves observed at low z (Hopkins et al. 2004), 
and it shows a very good agreement with the extinction curve predicted for dust formed
in SN ejecta. This is an indication that the properties of dust evolve
beyond $z \sim 6$.

Theoretical models, based on classical nucleation theory, predict that a few 
hundred days after the explosions silicate and carbon grains can form in 
expanding SN ejecta, with condensation efficiencies in the range $0.1-0.3$. 
This implies $\sim  0.1 - 1 M_{\odot}$ of dust per SN for stellar progenitor 
masses in the range of core-collapse SN ($12 - 40 M_{\odot}$) 
(Kozasa et al. 1991; Todini \& Ferrara 2001; Clayton,Deneault \& Meyer 2001; Nozawa et al. 2003).

There is now clear observational evidence for dust formation in core-collapse supernovae, but the quantity of dust formed within the
ejecta is still a subject of debate. The observed IR emission for a limited number of SNe, such as 1987A (Wooden et al. 1993),
1999em (Elmhamdi et al. 2003) and 2003gd (Sugerman et al. 2006; Meikle et al. 2007) implies dust masses which are 
generally smaller than $10^{-3} M_{\odot}$, corresponding to condensation efficiencies which are at 
least two orders of magnitude smaller than what theory predicts. Observations of young galactic SN remnants by 
the {\it Spitzer} telescope provide unambiguous evidence of dust formation 
(Hines et al. 2004). The inferred total mass of freshly formed dust is $0.02- 0.054 M_{\odot}$ for Cas A (Rho et al. 2008), 
and $0.015 M_{\odot}$ for 1E 0102.2-7219 (Rho et al. 2009). While masses up to $1-2 M_{\odot}$ have been estimated from FIR 
and submm observations (Dunne et al. 2003), the interpretation of these data is complicated by strong foreground 
contamination (Krause et al. 2004; Wilson \& Batrla 2005). The apparent discrepancies between observational estimates and 
theoretical models can be accomodated taking into account the partial destruction of newly synthesized 
dust in the reverse shock of the SN (Bianchi \& Schneider 2007).

The evolution of dust in the early Universe is still poorly understood. 
Models developed so far show that the current observational constraints 
on dust production in SNe are still consistent with the scenario where 
most of the dust in QSO at $z \sim  6$ is produced by SNe. 
However these estimates either neglect dust destruction 
by interstellar shocks (Maiolino et al. 2006) or make extreme 
assumptions on SN dust condensation factors (Dwek, Galliano \& Jones 2007). 

It is clear that our ability to interpret the observed properties of 
high redshift QSOs depends on a detailed understanding of the star 
formation history and the history of metal and dust pollution in their 
host galaxies. This, in turn, requires an adequate description of metal 
and dust stellar yields. 

In this work we study the evolution of dust produced by low, 
intermediate and high-mass stars as a function of time and metallicity, 
taking into account the stellar evolutionary timescales, i.e. without 
adopting an instantaneous recycling approximation. The aim of this study is to 
explore the relative importance of the two main stellar sources, AGB stars and 
SNe, in the evolution of dust at high redshift for galaxies with different star 
formation histories (SFH) and stellar Initial Mass Function (IMF). 

The paper is organized as follows. In section 2, we describe 
the adopted grid of dust yields for AGB stars and SNe with different 
initial metallicities. In section 3 we discuss the time evolution of the 
predicted dust mass for two different SFH. In this section we also
explore the dependence of the results on the adopted IMF and the 
stellar mass-lifetime relation. In section 4 we present a chemical 
evolution model which includes dust destruction by interstellar SN 
shocks and, in section 5, we apply it to the QSO SDSS J1148+5251 at $z = 6.4$, 
using the star formation history of the host galaxy predicted by the 
numerical simulation of Li et al. (2007). Finally, in section 6
we discuss and summarize the results of our work. 
In what follows, we assume a $\Lambda$CDM 
cosmology with $\Omega_m = 0.3$, $\Omega_\Lambda = 0.7$, $\Omega_b = 0.04$, 
and $H_0 = 70$~km/s/Mpc. The age of the Universe at a redshift $z = 6.4$ is 
about 840 Myr. 

\section{Stellar sources of dust}

Dust production in the ISM is extremely inefficient as the timescales for 
dust formation are longer than the destruction timescales (Tielens 1998). 
Thus, formation sites during stellar evolution are required to explain the 
vast amounts of dust observed in the Universe. 

Most of the theoretical models for dust formation are based on classical nucleation theory (Feder et al. 1966): the formation 
of solid materials from the gas phase can occur only from a vapour in a supersaturated state. Condensation of a given
species occurs when its partial pressure in the gas exceeds its vapour pressure in condensed phase (Whittet 1992), 
with subsequent particle growth by random encounters leading to cluster formation. 
Chemical reactions will further assist cluster production.
The rate of grain growth therefore depends on the temperature, T$_{g}$, and pressure, P$_{g}$, of the gas and the 
condensation temperature of a species, T$_{c}$. 
The optimum conditions for nucleation are thought to occur in the range  $10^{-8}\leq$P$_{g}$(Pa)$\leq 10^{5}$ and 
T$_{g}\leq 1800$~K (Salpeter 1974). This gives very stringent restrictions on where nucleation can occur. 

Stars seen to produce dust, only satisfy the conditions needed for the 
nucleation processes at the very latest stages of their evolution. 
The main stellar sites of dust grain condensation are the cool, dense 
atmospheres of AGB stars and the expanding SN ejecta. The type of dust 
formed depends on the relative abundance of carbon and oxygen at
the onset of nucleation.
 
In this work, we use the results obtained by Bianchi \& Schneider (2007) to estimate the mass and properties 
of dust formed in Type II SN ejecta and those presented by Zhukovska, Gail \& Trieloff (2008) to take into account grain 
condensation in the atmospheres of AGB stars. 

\begin{figure*}
\centering
\includegraphics[width=16.0cm]{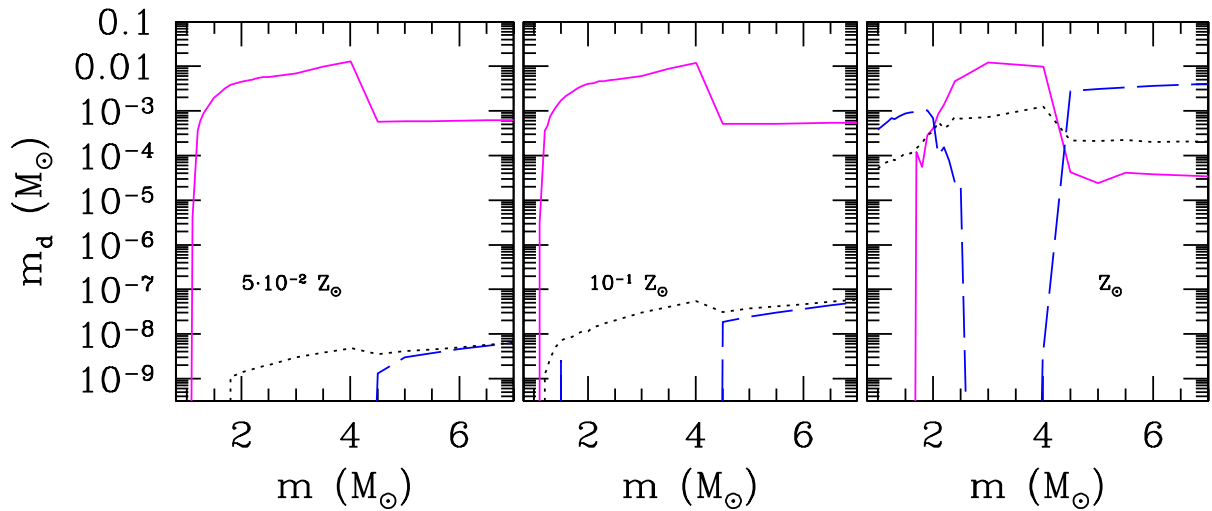}
\caption{Dust masses returned by AGB stars as a function of the initial stellar mass for three different
initial metallicities: $5 \times 10^{-2} Z_{\odot}$ (left panel) , $10^{-1} Z_{\odot}$ (central panel) and
$Z_{\odot}$ (right panel). In each panel, the dust mass is separated into three main components: carbon 
(solid line), silicate (dashed line), and other dust (the latter component here represents the sum of 
Fe and SiC dust, dotted line). The results are taken from the grid of models by Zhukovska et al. (2008).} 
\label{fig:dustAGB} 
\end{figure*}

\subsection{Dust produced by AGB stars}

All stars with initial masses $0.8 M_{\odot} \lesssim m \lesssim 8 M_{\odot}$ on the main sequence 
evolve through the AGB phase of stellar evolution shortly before they become white dwarfs.
In these stars, the sites of dust production are their cool, dense atmospheres during the thermally 
pulsating (TP) AGB phase. In this evolutionary stage, stars loose mass at an extremely high rate by a 
stellar wind and develop optically thick circumstellar dust shells by condensation of solid particulates 
in the outflowing gas. 

Depending on the relative abundances of carbon and oxygen (C/O) of the ejected matter in which grains are 
formed, different dust mixtures are produced. Ferrarotti \& Gail (2006) provide models which consistently 
describe the dependence of dust production by AGB stars on stellar initial mass and metallicity. 
They calculate the dust production rate of the dominant dust species, namely silicates, carbon, silicon carbide 
(SiC) and iron, which are formed from the most abundant elements. 
They use AGB stellar yields from van den Hoek \& Groenewegen (1997) and Karakas et al. (2003).

On the basis of this model calculations, Zhukovska et al. (2008) propose refined grids of the ejected mass 
of different dust species, for different star masses in the range $(1-7) M_{\odot }$ and at 
various (sub-solar and solar) metallicities.
These models predict that silicate dust is produced by stars with initial masses $< 1.5 M_{\odot}$ and $> 4 M_{\odot}$, 
which during the AGB phase are oxigen-rich. These stars are spectroscopically identified as M (characterized by C/O$<1$) 
and S type (C/O$\simeq 1$). On the other hand, carbon and SiC dust production is dominated by stars with initial mass 
$(1.5 - 4) M_{\odot}$, during the carbon-rich phase of evolution on the AGB (C stars, C/O $>1$). 
Iron dust should be formed in all M, S and C type stars. 
However, only some hints of its existence have been found up to now (e.g. Kemper et al. 2002).
 
To ease the comparison with the dust mass produced by SNe, in Fig. 1 we plot the mass of carbon, silicate and 
\emph{other} dust species produced by AGB stars as a function of the initial star mass and for three different 
metallicities, $(5 \times 10^{-2}, 10^{-1}, 1) Z_{\odot}$. Here we indicate as \emph{other} dust 
the sum of iron and SiC grains. 
It is clear from the figure that AGB stars can produce $10^{-5} < M_{\rm carb}/ M_{\odot} \leq 10^{-2}$ at 
all metallicities while comparable amounts of silicate can only be produced when $Z \sim Z_{\odot}$. In fact, silicate 
grain condensation requires the presence of a sufficient amount of Si, O, Mg and Fe in the stellar outflow. 
These metal species cannot be produced by the star and have to be collected by the pre-enriched ISM 
prior to star formation.

\begin{figure*}
\centering

\includegraphics[width=16.0cm]{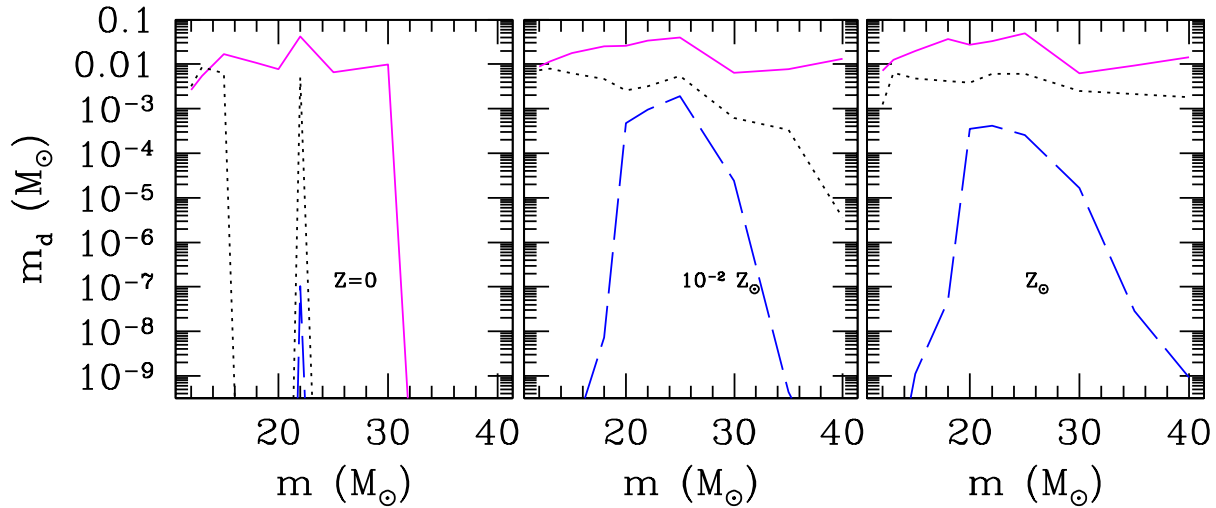}
\caption{Dust mass returned by supernovae as a function of the stellar progenitor mass and for three different 
initial metallicities, $0$ (left panel), $10^{-2} Z_{\odot}$ (central panel) and $Z_{\odot}$ (right panel). 
These models refer to an assumed ISM density of $\rho_{ISM}=10^{-24}$g cm$^{-3}$, when 7\% of the original dust 
mass survives the passage of the reverse shock. The three main dust species are compared: graphite (solid line), 
silicates (dashed line) and other species, which here represents the sum of Fe$_{3}$O$_{4}$ and Al$_{2}$O$_{3}$ grains 
(dotted line).}
\label{fig:dustSN} 
\end{figure*}

\subsection{Dust produced by SNe}

Stars with masses $m >8 M_{\odot}$ end their life exploding as core-collapse SNe. Main sites of dust formation are their 
expanding ejecta, a few hundred days after the explosions. Dust formed in stellar winds 
during the pre-supernova phase (i.e. by Red Supergiant stars or Wolf-Rayet stars) is almost entirely 
destroyed by the subsequent shock wave (see discussion in Zhukovska et al. 2008 and references therein).

Most of the available models proposed to investigate the process of dust formation in expanding SN ejecta
are based on classical nucleation theory and grain growth. The onset of grain formation depends on the 
temperature structure in the SN ejecta, whereas the grain composition mainly reflects its chemical composition, which 
depends on the nucleosynthesis occurring during stellar lifetime (i.e. on the progenitor mass) 
and explosion mechanism. Thus, models of dust formation in SN ejecta are based on specific 
prescriptions for the chemical composition and thermodynamics of the expanding gas. 

For the present analysis, we use the results of the model initially developed by Todini \& Ferrara (2001) 
but recently revisited by Bianchi \& Schneider (2007) in order to follow the evolution of newly condensed 
grains from the time of formation to their survival through the passage of the reverse shock in the SN remnant.
In these works, the ejecta are taken to have a uniform composition and density, with initial temperature and
density chosen to match the observations of SN1987A. The initial composition depends on the metallicity and
mass of the progenitor star while the dynamic is given by the mass of the ejecta and the kinetic energy of
the explosion: the initial SN models were taken from Woosley \& Weaver (1995). Therefore, we consider
dust formed in the ejected matter by SNe with initial progenitor star masses in the range $(12-40) M_{\odot}$ 
and metallicities $(0, 10^{-4}, 10^{-2}, 10^{-1}, 1) Z_{\odot}$.
Bianchi \& Schneider (2007) derive the fraction of dust which survives the passage of the reverse 
shock, as a function of progenitor star mass and metallicity. Dust destruction due to the passage of the reverse shock depends 
on the density of the surrounding ISM. The denser is the gas in the ISM, the higher is the fraction of the destroyed dust mass. 
We refer the reader to the original paper for more details on the model. 

In what follows, we adopt the results of the reference model by Bianchi \& Schneider (2007), i.e. the model which starts
grain condensations from seed clusters made of $N \geq 2$ monomers, assumes a sticking coefficient $\alpha=1$ 
(all the gas particles colliding on a grain will stick to it) and an ISM density of $\rho_{\rm ISM}=10^{-24}$g cm$^{-3}$. 
In this model, about $0.1-0.6 M_{\odot}$ of dust are formed in the ejecta but only 7\% of the initial dust mass survives 
the passage of the reverse shock. 
Note, however, that these numbers are sensitive to variations on the assumed values of $N$, $\alpha$ and $\rho_{\rm ISM}$. 
For a larger $N$ and/or a smaller $\alpha$, dust grains are characterized by smaller radii, which are more easily destroyed. 
If the SN explodes in a denser ( $\rho_{\rm ISM}>10^{-23}$g cm$^{-3}$) ISM, only about 2\% of the dust mass survives.
Conversely, for a lower density ISM ($\rho_{\rm ISM}=10^{-25}$g cm$^{-3}$) a larger fraction, about 20\%, is left.

In Fig.~2 we present the final mass of carbon, silicate and other dust species (essentially magnetite, Fe$_3$O$_4$, and 
corundum, Al$_2$O$_3$) for the fiducial model, as a function of the initial stellar mass and metallicity, as given by 
Bianchi \& Schneider (2007). Carbon dust condensation efficiency does not vary much with metallicity, with
$10^{-3} \leq M_{\rm carb}/M_{\odot} \leq 10^{-1}$, while silicates 
production appears to be inefficient at low metallicity. 
It is worth noting that the model by Bianchi \& Schneider (2007) shows that 
SNe are expected to produce mainly carbon, magnetite and corundum grains. 
This is at odd with the models developed by Nozawa et al. (2003, 2007),
in which the composition is dominated by silicate and iron grains.
The differences are likely to be due to different initial SN models and their
assumption of complete formation of SiO and CO  molecules. While the first 
molecule is necessary to study the formation of Si-bearing grains, the 
second may be a sink for carbon atoms that otherwise would accrete on grains 
(Bianchi \& Schneider 2007). In both models, though, a similar amount of dust 
is formed and survives the reverse shock.  A thorough comparison between 
different theoretical models is deferred to future studies.

Finally, it should be mentioned that in the present model we do not consider the
possible contribution to dust production by pair-instability and type Ia SNe. 
The former class is predicted to be the end-product of the evolution of very massive 
Population III stars with masses in the range $140 M_{\odot} \leq M \leq 260 M_{\odot}$ and
might provide an important source of dust at high redshift (Schneider, Ferrara \& Salvaterra 2004;
Nozawa et al. 2003). However, our aim here is to compare the relative contribution of the two
main stellar sources of dust assuming a conventional stellar initial mass function with an
upper mass cut-off of $100 M_{\odot}$. For the latter class, there are no available theoretical
models for dust production and destruction, although Clayton et al. (1997) 
have presented arguments in favor of dust production in type Ia SNe, with important
consequences for the composition of iron dust in the interstellar medium (Dwek 1998).   
Type Ia SNe might be relevant for dust evolution at high redshift given the recent observational
evidence for a "prompt" component exploding in a time scale of
the order of 100 Myr after the stellar birth (Mannucci, Della Valle \& Panagia 2006) and that
a sizeable fraction of stars with masses in the range $3 - 8 M_{\odot}$ is predicted to
explode as type Ia SNe (Maoz 2008). However, adding the contribution of 
type Ia SNe to (iron) dust production is beyond the scope of the present analysis.

\section{Cosmic dust yields}

As a first step of the analysis, we compute the time evolution of  
the total dust mass produced, $M_{d}(t)$, considering the contribution 
of AGB stars and Type II SNe discussed above.
The aim of this study is to investigate the relative importance of these 
two formation channels and the characteristic timescales at which one 
dominates over the other.

We take into account that stars of different masses evolve on different 
timescales\footnote{The characteristic times for dust formation in AGB and dust formation/evolution in SNe are negligible compared to the star 
lifetime (Bianchi \& Schneider 2007).}. Therefore, the time evolution of the
total dust mass can be expressed as:

\begin{equation}
 M_d (t) = \int_{0}^{t} dt' \int_{m_{\ast}(t')}^{100 M_{\odot}} m_{d}(m) \phi(m) \rmn{SFR}(t'-\tau_{m}) dm,
\end{equation}
\noindent
where $m$ is the progenitor stellar mass, $m_{d}$ is the dust mass produced by a star of mass $m$, 
$\phi(m)$ is the stellar Initial Mass Function (IMF), $\rmn{SFR}$ is the star formation rate, and $\tau_m$
is the stellar lifetime of a star of mass $m$. 
The lower limit $m_{\ast}$ is the minimum progenitor mass contributing to the dust production at a time 
$t'$, i.e. the mass corresponding to a stellar lifetime $\tau_{m}=t'$.

Note that we have no prescriptions for dust production by stars in the mass range $(8-11)$ M$_{\odot}$, which
evolve as AGB and finally explode as core-collapse SNe (II-P SNe, Smartt et al. 2009).
Dust yields for stars in this mass range (and for different metallicity) are not yet available in the literature. 
Therefore we extrapolate from the dust mass produced by the largest available AGB model (7 $M_{\odot}$) and by the smallest
SN progenitor (12 $M_{\odot}$) to account for metal and dust production in this mass range. 
However, the contribution to the total mass of dust from stars in this mass range is less than 5\%).

To investigate the dependence on the star formation history, we make two extreme hypotheses:  
(i) a burst occurring at $t=0$, where all stars are born at the same epoch, in an instantaneous episode of star 
formation; (ii) a constant $\rmn{SFR}$, where stars form continuously and at a constant rate for 10~Gyr.

It is clear that the relative contribution of AGB stars and SNe depends on the shape of the stellar IMF and
on the adopted stellar lifetimes.
In our reference model, we assume that stars form with masses in the range $(0.1 - 100)$ M$_{\odot}$ according
to a Larson IMF (Larson 1998), which follows a Salpeter-like power law at the upper end but flattens below a 
characteristic stellar mass:
\begin{equation}
\phi(m) \propto m^{-(\alpha + 1)} e^{-m_{ch}/m},
\label{imf}
\end{equation}
\noindent
where $\alpha=1.35$, $m_{ch}=0.35 M_{\odot}$ and we normalize the integral of $m\phi(m)$ in the mass range $(0.1-100)$ M$_{\odot}$ to unity.
The lifetimes of stars of different mass and metallicity are computed according to the following
simple parametric form proposed by Raiteri et al. (1996),
\begin{equation}
\log (\tau_{m}) = a_{0}(Z) + a_{1}(Z) \log m + a_{2}(Z) (\log m)^{2}, 
\end{equation}
\noindent
where $\tau_{m}$ is expressed in yr, $m$ in solar units and the coefficients are:
\begin{eqnarray}
&&a_{0}(Z) = 10.13 + 0.07547 \log Z - 0.008084 (\log Z)^{2}, \\  
&&a_{1}(Z) = -4.424 - 0.7939 \log Z - 0.1187(\log Z)^{2}, \\ 
&&a_{2}(Z) = 1.262 + 0.3385 \log Z + 0.05417 (\log Z)^{2}. \nonumber
\end{eqnarray}
\noindent
Although this relation has been tested to reproduce the stellar evolutionary data in the
metallicity range  $3.5 \times 10^{-3} \leq Z/Z_{\odot} \leq 1.5$, we apply it also to stars
with lower metallicities.

\begin{figure}
\centering
\includegraphics[width=9.0cm]{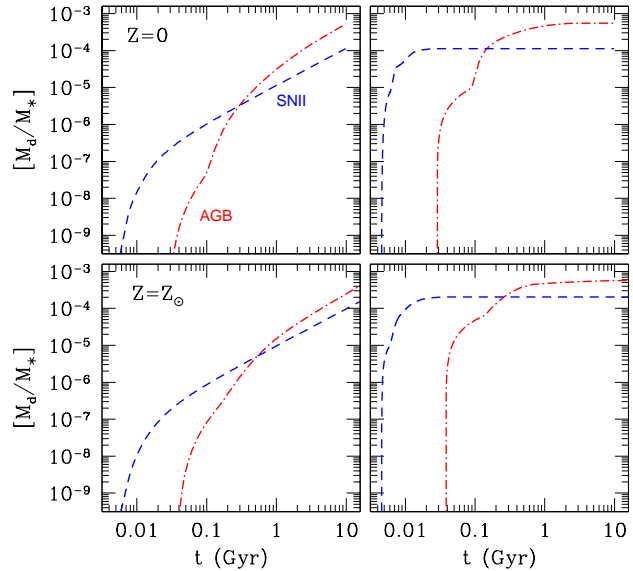}
\caption{Time evolution of the mass of dust produced by SNe (dashed lines) and AGB stars (dot-dashed lines) normalized to the
total mass of stars formed in $10$~Gyr in our reference model (see text). 
All the stars have an initial metallicity of $Z=0$ (upper panels) and $Z=Z_{\odot}$
(lower panels). The rigth panels show the results obtained assuming that star formation occurs in a burst at $t=0$ whereas the left panels
are computed for a constant star formation rate lasting for 10~Gyr.}
\label{fig:mdust_ref}
\end{figure}

In Fig.~\ref{fig:mdust_ref} we show the time evolution of the dust mass normalized to the final mass of stars 
for a burst-like (right panels) and a constant star formation history (left panels) and assuming that 
all stars form at the same initial metallicity of 
$Z=0$ (upper panels)\footnote{As the grids of Zhukovska et al. (2008) are not computed for 
$Z <5 \times 10^{-2} Z_{\odot}$, we have assumed that the AGB dust mass for lower metallicities 
is the same as at $Z =5 \times 10^{-2} Z_{\odot}$. For both AGB and SN dust, dust mass values at 
metallicities not available in the grids are obtained through interpolation.}
 (upper panels) and $Z=Z_{\odot}$ (lower panels).
The curves represent the mass of dust, summed over all the components (carbon, silicates and other
dust species) produced by SNe (dashed lines) and AGB stars (dot-dashed lines).
At the beginning of the evolution, dust production is dominated by SNe; however, for $t > 30$~Myr
the contribution of AGB stars rapidly increases and becomes dominant in a characteristic time 
which depends both on the assumed star formation history and stellar metallicity, ranging
between 150~Myr (for a burst at $Z = 0$) and 500~Myr (for a constant star formation rate at $Z = Z_{\odot}$). 
These results suggest that AGB stars might give a non-negligible contribution to dust production 
at early cosmic epochs and that their role as dust factories at $z \geq 6$ might have been over-looked.

We have also investigated how these results depend on different model assumptions, 
such as the stellar lifetime relation and IMF. It can be shown that the timescales 
at which AGB stars start to be the dominant dust sources change by less than 20\% if we adopt the stellar 
lifetimes proposed by Padovani \& Matteucci (1993), which predicts slightly longer 
lifetimes for stars of intermediate mass and high metallicity compared to our reference model. 

\begin{figure}
\centering
\includegraphics[width=8.5cm]{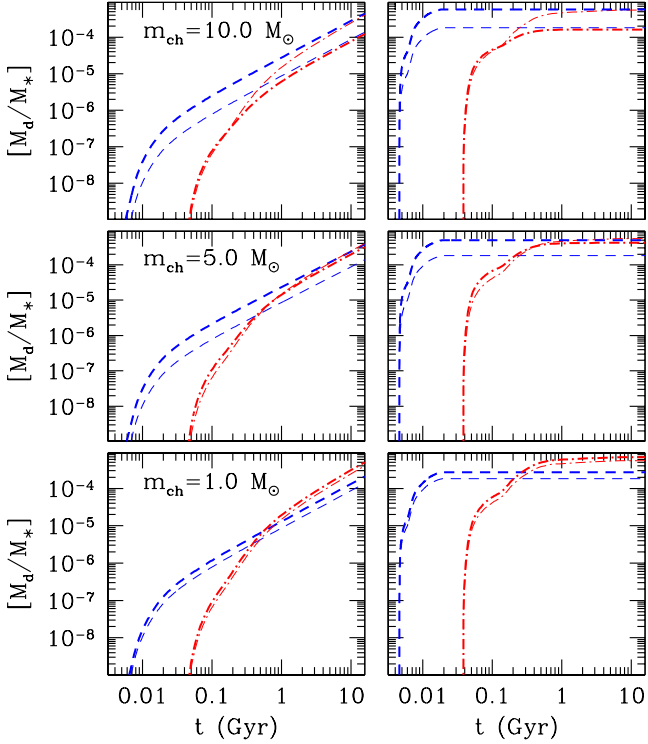}
\caption{The same as in Fig.\ref{fig:mdust_ref} but assuming that all stars form with a metallicity $Z=Z_{\odot}$
and a Larson IMF with characteristic masses $1$, $5$, and $10 M_{\odot}$ (from bottom to top); left panels are
computed assuming a constant star formation rate for 10~Gyr; in right panels all stars are formed in a burst at $t=0$.
The mass of dust is normalized to the final stellar mass formed. For comparison, in each panel we also show
the results of the fiducial model of Fig.~\ref{fig:mdust_ref} with thin lines.} 
\label{fig:mdust_mch}
\end{figure}

The largest uncertainty is introduced by the shape of the stellar IMF, which can be quantified through 
variations of the stellar characteristic mass $m_{ch}$ in eq.~\ref{imf}. In Fig.~\ref{fig:mdust_mch} 
we show the time evolution of dust mass produced by SNe and AGB stars assuming a 
constant stellar metallicity of $Z=Z_{\odot}$ and three different values for the 
characteristic mass $m_{ch} = 1, 5,$ and $10 M_{\odot}$ (from bottom to top).
As in Fig.~3, stars are formed at a constant rate for $10$~Gyr (left panels) and in a burst at $t=0$ (right
panels) and the resulting dust mass is normalized to the total stellar mass formed. For comparison, in 
Fig.~\ref{fig:mdust_mch} the corresponding reference models are also shown (light curves). 

As it can be inferred from the figure, the evolution is almost identical to the reference model for $m_{ch} = 1 M_{\odot}$;
for $m_{ch} = 5 M_{\odot}$, the mass of dust contributed by SNe is increased (as a 
consequence of the larger SNe rate) and AGB stars produce a comparable amount of dust on timescales $\sim 1$~Gyr.
Finally, when $m_{ch} = 10 M_{\odot}$ the AGB dust component is always sub-dominant, as a result of the larger
SN rate and of the smaller number of stars formed with intermediate masses. Thus, it is clear that variations
of the stellar characteristic mass over cosmic history due to the smaller gas metallicity and/or to the larger 
cosmic microwave background temperature (see, e.g., Omukai et al. 2005) might significantly affect the origin
and properties of dust in the high redshift Universe.
 
\section{Chemical evolution model with dust}

In this section, we describe a simple chemical evolution model with dust which allows to follow the 
time evolution of the mass of gas, stars, metals in specific environments, such as high redshift galaxies 
and quasars, given the appropriate star formation history. 
We then apply this model to a ``template'' QSO host galaxy at high redshift. 

In this model, we make the simplifying assumptions that (i) there are no mass exchanges with the surrounding 
intergalactic medium (IGM) (``close-box'' model), and that (ii) dust produced by AGB stars and SNe does not 
grow in molecular clouds (MC) by accretion of metals. Indeed, dust grains formed in stellar outflows can serve 
as growth centers for accretion of icy mantels in dense MC (e.g. Draine 1990). However, grain mantles are 
amorphous and heterogeneous, thus MC-grown dust has different properties from 
dust produced by stars and must be treated as a separate component (see e.g. Zhukovska et al. 2008 and 
references therein). Since we are primarily interested in the study of the relative contribution of the two 
main stellar sources, we do not follow the evolution of this component.

The model consists of a set of differential equations regulating the time evolution of the different components: 
\begin{eqnarray}
\frac{dM_{\ast}(t)}{dt} & = & \rmn{SFR}(t) - \frac{dR(t)}{dt}\\
\frac{dM_{\rm ISM}(t)}{dt} & = & -\rmn{SFR}(t) + \frac{dR(t)}{dt} \\
\frac{dM_{Z}(t)}{dt} & = & -Z_{\rm ISM}(t) \rmn{SFR}(t) + \frac{dY_{Z}(t)}{dt} \\
\frac{dM_{d}(t)}{dt} & = & -Z_{d}(t) \rmn{SFR}(t) + \frac{dY_{d}(t)}{dt} -\frac{M_{d}(t)}{\tau_{d}} \label{eq:dust}
\end{eqnarray}
where $M_{\ast}$ and $M_d$ are the total masses of stars and dust, $\rmn{SFR}$ is the star formation rate.
Since a fraction of dust can be destroyed by interstellar SN shocks and returns metals back into the gas phase, 
we avoid complications indicating with $M_{\rm ISM}$ the total mass in the interstellar medium 
(gas and dust) and with $M_Z$ the total mass of metals (diffused in the gas phase and condensed into dust grains). 
At any given time, the mass 
of {\it gas} can be computed as the difference between $M_{\rm ISM}$ and $M_d$, and the mass of {\it gas phase} metals as the difference between
$M_Z$ and $M_d$ (with proper stoichiometric coefficients when we are interested in a particular gas phase element).
The quantities $Z_{\rm ISM}(t)=M_{Z}(t)/M_{\rm ISM}(t)$ and $Z_{d}(t)=M_{d}(t)/M_{\rm ISM}(t)$ are the total metal and dust abundances 
in the ISM. 
The terms $dR/dt$, $dY_{Z}/dt$ and $dY_{d}/dt$ represent the rates at which mass (gas and dust), heavy elements (gas-phase and condensed) and dust are 
returned to the ISM. These terms are given by:
\begin{eqnarray}
\frac{dR(t)}{dt} & = & \int_{m_{\ast}(t)}^{100 M_ {\odot}} (m - \omega_{m}(m, Z_{\rm ISM})) \phi(m) \rmn{SFR}(t-\tau_{m}) dm \\
\frac{dY_{Z}(t)}{dt} & = & \int_{m_{\ast}(t)}^{100 M_ {\odot}} m_{Z}(m, Z_{\rm ISM}) \phi(m) \rmn{SFR}(t-\tau_{m}) dm \\
\frac{dY_{d}(t)}{dt} & = & \int_{m_{\ast}(t)}^{100 M_ {\odot}} m_{d}(m, Z_{\rm ISM}) \phi(m) \rmn{SFR}(t-\tau_{m}) dm \label{eq:ejerate}
\end{eqnarray}
where $m$ is the initial mass of the progenitor star, and the lower limit $m_{\ast}$ is the mass with a
lifetime $\tau_{m}=t$. 
The total mass of metals (pre-existing and newly synthesized) produced by a star of initial mass $m$ and metallicity $Z$, $m_{Z}(m,Z)$, and the mass of the stellar
remnant $\omega_m(m, Z)$, are taken from van den Hoeck \& Groenewegen (1997) for AGB stars with initial metallicities 
$Z=(5 \times 10^{-2}, 0.2, 1) Z_{\odot}$ and masses $1-8 M_{\odot}$ and from Woosley \& Weaver (1995) for SNe with initial masses 
$12 - 40 M_{\odot}$ metallicities  $Z=(0, 10^{-4}, 10^{-2}, 10^{-1}, 1)$ Z$_{\odot}$. 
The grid of dust masses, $m_{d}(m,Z)$, is the same as described in section 2 and shown in Figs.~1 and 2.
Note that integrating eq.~\ref{eq:ejerate} over time we obtain the mass of dust as given in eq. 1.
The last term of eq.~\ref{eq:dust} is the dust destruction rate in the ISM by thermal sputtering in high-velocity ($v>150$ km s$^{-1}$) SN shocks. 
The destructive effect of the SN shock waves on dust grains depends on the dust species and on the destruction timescale, $\tau_{d}$, i.e. 
the lifetime of the dust grains against destruction by SN remnant (Jones, Tielens \& Hollenbach 1996).
Following Dwek, Galliano \& Jones (2007) we assume that
\begin{equation}
\tau_{d}=\frac{M_{g}(t)}{m_{\rm ISM}R_{\rm SN}},
\end{equation}
\noindent
where $m_{\rm ISM}$ is the effective ISM mass that is completely cleared of dust by a single SN remnant (i.e. a measure of the grain destruction
efficiency), which depends on the dust species, and  $R_{\rm SN}$ is the SN rate,
\begin{equation}
 R_{SN}(t) = \int_{m >8 M_{\odot}}^{40 M_{\odot}} \phi(m) \rmn{SFR}(t-\tau_{m}) dm.
\end{equation} 
\noindent
Grain destruction efficiencies by radiative shocks have been modeled by Jones et al. (1994), and typical grain lifetimes 
of 0.6 and 0.4 Gyr have been estimated, respectively, for carbonaceous and silicate grains in the Milky Way (Jones et al. 1996).
Dwek, Galliano \& Jones (2007) investigate the dependence of grain destruction efficiency, $m_{\rm ISM}$, on the ISM density, 
both for silicate and carbon dust. Following their study, for a uniform ISM with density $\rho_{\rm ISM}=10^{-24}$g cm$^{-3}$, 
we assume $m_{\rm ISM} = 1200 M_{\odot}$ for silicates and $800 M_{\odot}$ for carbon and other dust species.
Finally, in what follows we always consider our fiducial model, i.e. a Larson IMF with a characteristic mass of 
0.35 $M_{\odot}$ and stellar lifetimes from Raiteri et al. (1996) as described in section~3.

\section{The most extreme QSO at redshift 6.4}

We apply the model described above to the most distant quasar, SDSS J1148+5251, discovered at redshift $z \simeq 6.4$ (Fan et al. 2003).
It is a very luminous quasar powered by a super massive black hole (SMBH) with mass $(1-5) \times 10^{9} M_{\odot}$ accreting 
close to its Eddington limit (Willott et al. 2003). The FIR luminosity is $L_{FIR} \sim 10^{13} L_{\odot}$, suggesting a mass of dust in excess of 
$10^{8} M_{\odot}$ (Bertoldi et al. 2003; Robson et al. 2004; Carilli et al. 2004; Beelen et al. 2006). If all the FIR emission is assumed to come
from young star, the derived $\rmn{SFR}$ is about $3 \times 10^{3} M_{\odot}$ yr$^{-1}$ (Bertoldi et al. 2003). Such a high $\rmn{SFR}$ is also 
supported by the first 
detection of the carbon [CII] line at 158 $\rm{\mu}$m (Maiolino et al. 2005). Alternatively, if the active galactic nucleus (AGN) contributes to
dust heating, the inferred $\rmn{SFR}$ would be much lower, in agreement with what implied by a simple application of the Schmidt-Kennicutt law
adopting a total gas mass of $\sim 10^{10} M_{\odot}$ (Walter et al. 2004), which yields a value of $200 M_{\odot}$ yr$^{-1}$ (Dwek, Galliano \& Jones 2007; 
Li et al. 2008).
  
The interpretation of the observed properties of J1148+5251 and of its host galaxy is therefore contradictory. Our aim here is to explore the
relative importance of SNe and AGB in the build-up of the dust mass inferred for this system. We adopt the star formation
history of the host galaxy of J1148+5251 predicted by a recent numerical simulation by Li et al. (2007). 
We then apply our chemical evolution model with dust and discuss the dependence of the results on the adopted star formation history. 
Note that this approach is not entirely self-consistent given that, as we will see below, the simulated system is not a closed-box.
A more accurate study of the chemical properties of the host galaxy of J1148+5251 implementing dust evolution in the semi-analytical merger-tree 
code GAMETE (Salvadori, Ferrara \& Schneider 2008) is currently underway. 

\begin{figure}
\centering
\includegraphics[width=8.0cm]{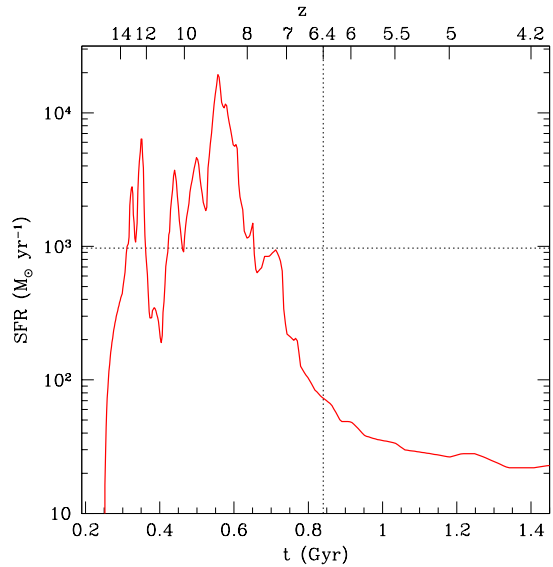}
\caption{Time evolution of the star formation rate in the host galaxy of SDSS J1148+5251 as predicted by the simulation of Li et al. (2007). 
The rapid bursts of star formation are determined by strong successive interactions between the progenitor galaxies during a period of time ranging from 
about 0.29 Gyr (the time at which structure formation is set to begin) and 0.84 Gyr (the Hubble time at the quasar redshift $z=6.4$). 
The $\rmn{SFR}$ is expressed in M$_{\odot}$ yr$^{-1}$ and the time in Gyr.
The horizontal line shows the average value of the star formation rate over the lifetime of the simulated QSO host galaxy, found to be $\sim 1000 M_{\odot}$ yr$^{-1}$,
while the vertical line indicates the age of Universe ($840$~Myr) corresponding to a redshift $z=6.4$, at which the quasar is observed.
}
\label{fig:Lisfr}
\end{figure}

The numerical simulation by Li et al. (2007) follows the hierarchical assembly of the SDSS J1148+5251 halo in a $\rm{\Lambda}$CDM cosmology. 
Their model accounts for the quasar activity, the host galaxy properties and includes a self-regulated growth of the SMBH. The quasar host galaxy is found to
form through seven major mergers of gas rich starburst progenitors between $z = 14.4$ and $6.4$. 
Intense starbursts are triggered by the gravitational interactions between the merging galaxies and rapid black hole accretion is fuelled by the highly 
concentrated gas (see Li et al. 2007 for details). This model is able to reproduce most of the observed properties of SDSS J1148+5251, such as the black 
hole mass, the quasar luminosity and metallicity. For convenience, Fig. \ref{fig:Lisfr} shows  the time evolution of the star formation rate found 
by Li et al. (2007). The star formation rate ranges between $\sim 100 M_{\odot}$ yr$^{-1}$, both at the onset of the interactions between the 
progenitor galaxies (at about 0.29~Gyr, corresponding to a redshift $z \sim 14$) and at the time of the final 
coalescence (0.8~Gyr, corresponding to $z \sim 6.5$), to values as large as $\sim 2 \times 10^{4} M_{\odot}$ yr$^{-1}$ at the 
final major merger, represented by the peak at 0.6 Gyr ($z\sim 8.5$). 
The horizontal line shows the average value of the star formation rate over the lifetime of the simulated QSO 
host galaxy, found to be $\sim 1000 M_{\odot}$ yr$^{-1}$.
Note that the simulation predicts a rate of about 100 M$_{\odot}$ yr$^{-1}$ at 840 Myr (corresponding to redshift $z=6.4$), an order of magnitude smaller 
than that inferred by FIR observations. Therefore, Li et al. (2008) suggests that the FIR luminosity of the QSO is not dominated by young stars but instead 
by the AGN (for over 80\%). 
Finally, the stellar mass of the quasar host at $z=6.5$ is predicted to be 
$\sim 10^{12} M_{\odot}$, in agreement with what would be required by the 
observed SMBH mass if the local black-hole-bulge relation were to hold at 
these high-redshifts. However, this result cannot be reconciled with 
observations of molecular gas: while the dynamical mass estimated from CO 
observations within 2.5 kpc from the QSO center is $\sim 5 \times 10^{10} 
M_{\odot}$, a conservative estimate of the contribution of the predicted 
stellar bulge within this region is a few $\times 10^{11} M_{\odot}$, 
about an order of magnitude larger (Walter et al. 2004). 

Using the simulated star formation rate evolution as an input to the chemical evolution model presented above, we can 
compute the corresponding evolution of the mass of stars and dust.  
Following Li et al. (2007), we assume that the QSO is hosted by a dark matter halo of mass $10^{13} M_{\odot}$ and that
the initial mass in baryons is equal to the universal value, $\Omega_b/\Omega_m$, i.e. $\sim 1.3 \times 10^{12} M_{\odot}$.

\begin{figure*}
\centering
\includegraphics[width=8.5cm]{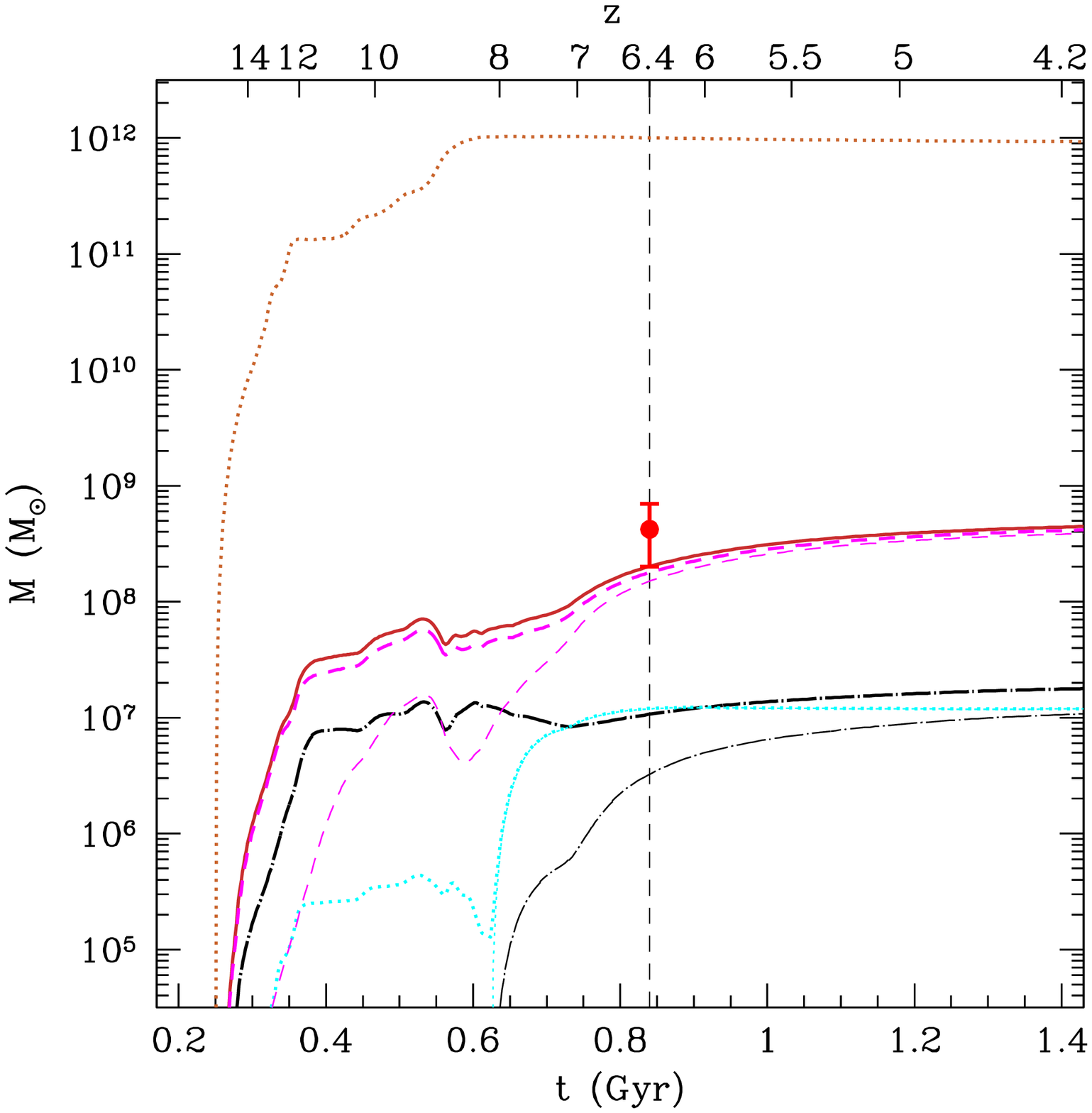}
\hspace{2mm}%
\centering
\includegraphics[width=7.35cm]{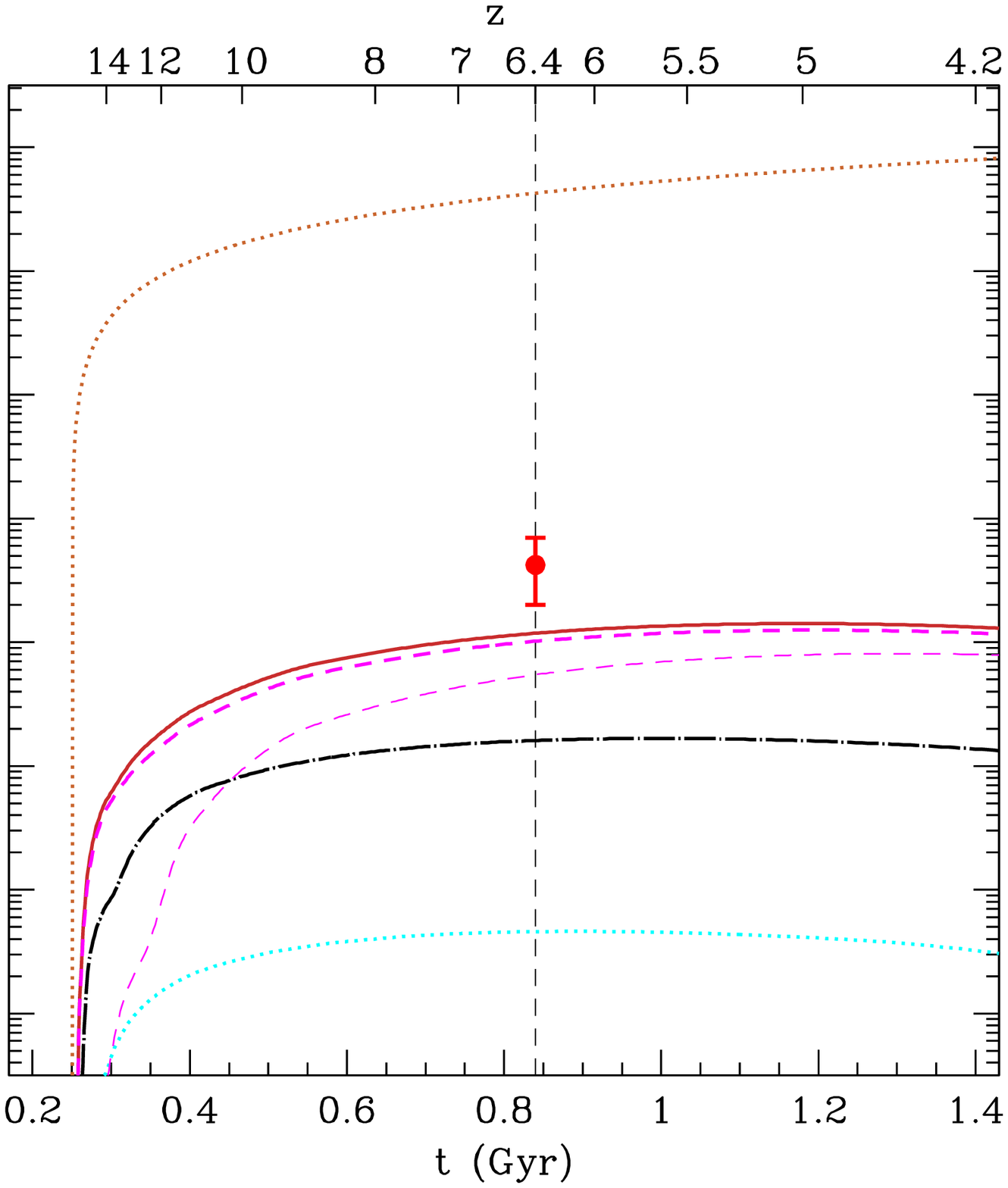}
\caption{Chemical evolution of the high redshift quasar SDSS J1148+5251 corresponding to the star formation histories shown in Fig.~\ref{fig:Lisfr}
(left panel, $\rmn{SFR}$ simulated by Li et al. (2007); right panel, constant $\rmn{SFR}$ with the average value of $~1000 M_{\odot}$yr$^{-1}$). 
The lines represent the evolution of the mass of stars (dotted) and dust 
(solid); the lower dashed, dotted and dashed-dotted lines indicate the time evolution of the three main dust
components, carbon, silicate and other dust grains, respectively. Thin lines represent the contribution of AGB stars only.
The vertical dashed line indicates the Hubble time corresponding to $z=6.4$ and the point with errorbars the assumed value for the
observed dust mass (see text).}
\label{fig:Liqso}
\end{figure*}

In the left panel of Fig.~\ref{fig:Liqso} we show the evolution of the mass of stars (dotted line), and dust (solid line) 
of the host galaxy. The vertical solid lines indicate the age of Universe at $z=6.4$ when the quasar is observed ($840$~Myr) 
and the data point represents the dust mass inferred from mm/submm observations by Beelen et al. (2006), 
$M_{\rm dust} = 4.2 \times 10^8 M_{\odot}$, with error-bars which account for the range of values quoted in the 
literature, $(2-7)\times 10^8 M_{\odot}$. The lower dashed, dotted and dashed-dotted lines indicate the time evolution of the three main dust
components, carbon, silicate and other dust grains, respectively. For these components, thin lines represent 
the contribution of AGB stars only.  

The right panel shows the same quantities computed assuming a constant star formation rate equal to $1000 M_{\odot}$yr$^{-1}$ 
(the average value shown with the horizontal line in Fig.~\ref{fig:Lisfr}). In this model
the total mass of stars formed at $z = 6.4$ is  $\sim 4.3 \times 10^{11} M_{\odot}$, marginally consistent with the 
observed dynamical mass after correcting for the extension of the stellar bulge.  

It is clear from the figure that our chemical evolution model with the simulated $\rmn{SFR}$ predicts a total mass of dust
of $M_{\rm dust} = 2 \times 10^8 M_{\odot}$ at $z =6.4$ in agreement with observations. We stress that this result is obtained
following dust production by stellar sources on their characteristic stellar lifetimes and dust destruction by astration and SN shocks.
The latter mechanisms are responsible for the down-turns in the evolution of the dust masses shown in the figure. 
The amount of dust destroyed by interstellar SN shocks depends on the age of the system: at the time corresponding to
$z = 6.4$, we find that this effect reduces the mass of dust by 30\% in both star formation models. On longer timescales,
dust destruction is more pronounced in the case of a constant star formation rate (SN progenitors continue to form) resulting
in more than 50\% dust mass reduction at $z = 4.1$. Conversely, assuming the simulated star formation history, the fraction of
dust destroyed at $z = 4.1$ is only 20\%. This is due to SFR (and therefore SN rate) quenching by AGN-feedback and by the larger 
relative contribution of AGB stars in this model (see below). 

As expected on the basis of the adopted stellar dust yields, the dominant dust component is found to be carbon dust, 
with silicates and other dust species being a factor 10 less abundant. 
Moreover, AGB stars are found to give a significant contribution to dust production at 
high redshift, accounting for $\sim$80 \% of the total mass of dust at $z = 6.4$. 
In particular, AGB stars start to contribute to carbon dust production already 200 Myr after the onset of star formation, 
but their contribution remains sub-dominant until the last major burst of star formation, which occurs at $z=8.5$ (0.6 Gyr). 
Following this burst, the mass of dust produced by AGB stars steeply rises for all the dust species. In fact, stars formed 
in this last major burst have nearly solar metallicities and their contribution to silicates and other dust components is 
no longer negligible.   

These results depends on the adopted star formation history. When a constant star formation rate is assumed, as shown in the
right panel of Fig. \ref{fig:Liqso}, the total mass of dust at $z = 6.42$ is a factor 2 smaller and the contribution of AGB
stars is $\sim$50 \%. Since the stellar metallicities in this case are smaller, the contribution of AGB stars to silicate and other
dust components are negligible. 

We stress that, although the star formation history of the QSO host galaxies is still largely unknown, our analysis indicate that 
the contribution of AGB stars to dust production cannot be neglected, even at the most extreme redshifts currently accessible to
observations. 
 
Additional constraints on the chemical evolution model can be obtained from observations of the metallicity and elemental abundances in the 
Broad Line Region (BLR) of the QSO. In fact, the detection of metal lines such as Fe (Barth et al. 2003), CII (Maiolino et al. 2005), and OI 
(in absorption, Becker et al. 2006) indicates that the ISM of the QSO host was significantly enriched by heavy elements.  
In a recent analysis, Juarez et al. (2009) investigated the metallicity of the BLR of a sample of 30 QSOs in the range 
$4 < z < 6.4$, including J1148+5251, using the (SiIV+OIV)/CIV ratio, which is a good metallicity tracer (Nagao et al. 2006). 
According to this analysis, the observed ratios do not show any correlation with redshift and correspond to metallicities 
which are several times solar. 

\begin{figure*}
\centering
\includegraphics[width=8.5cm]{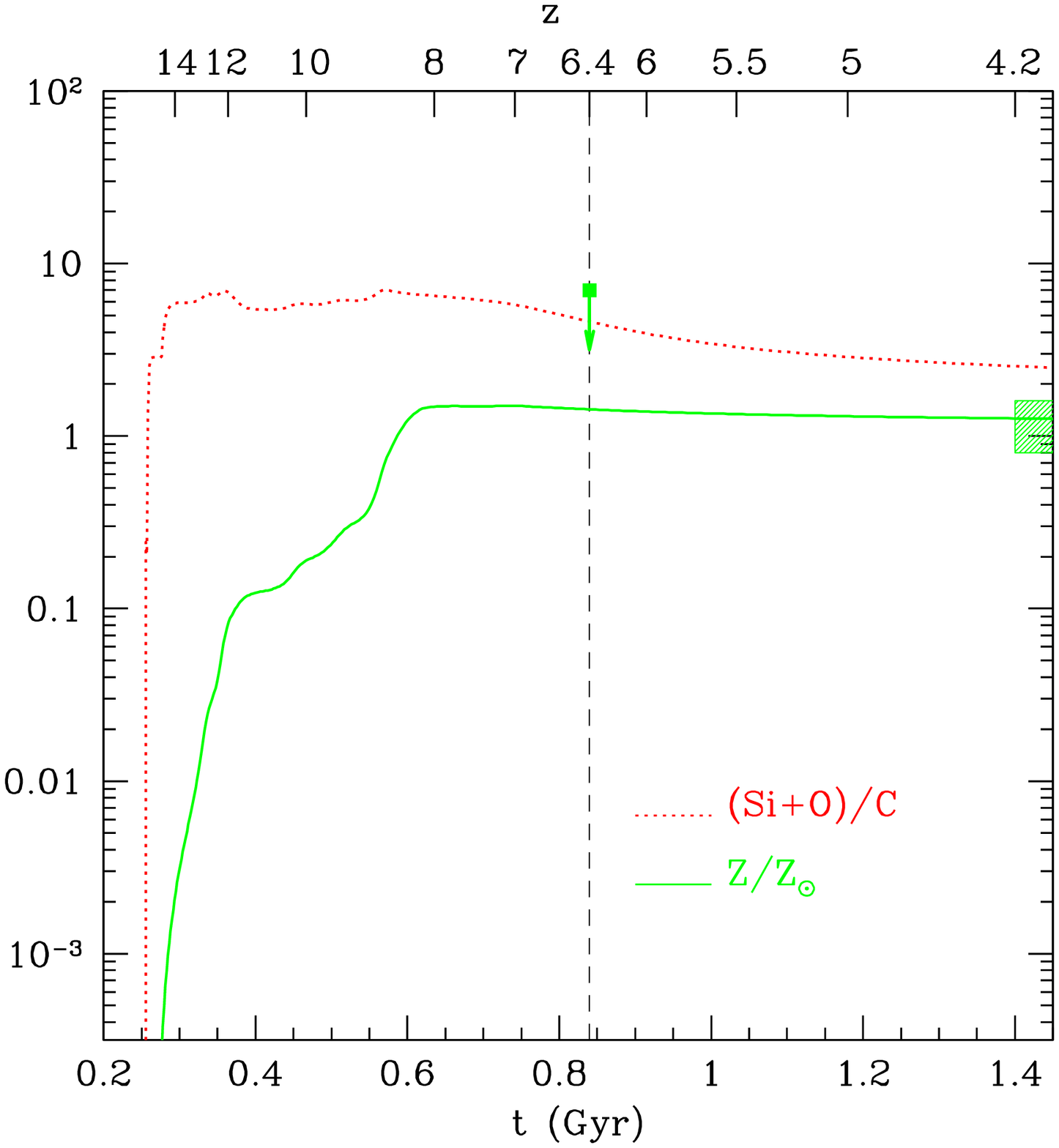}
\centering
\includegraphics[width=7.9cm]{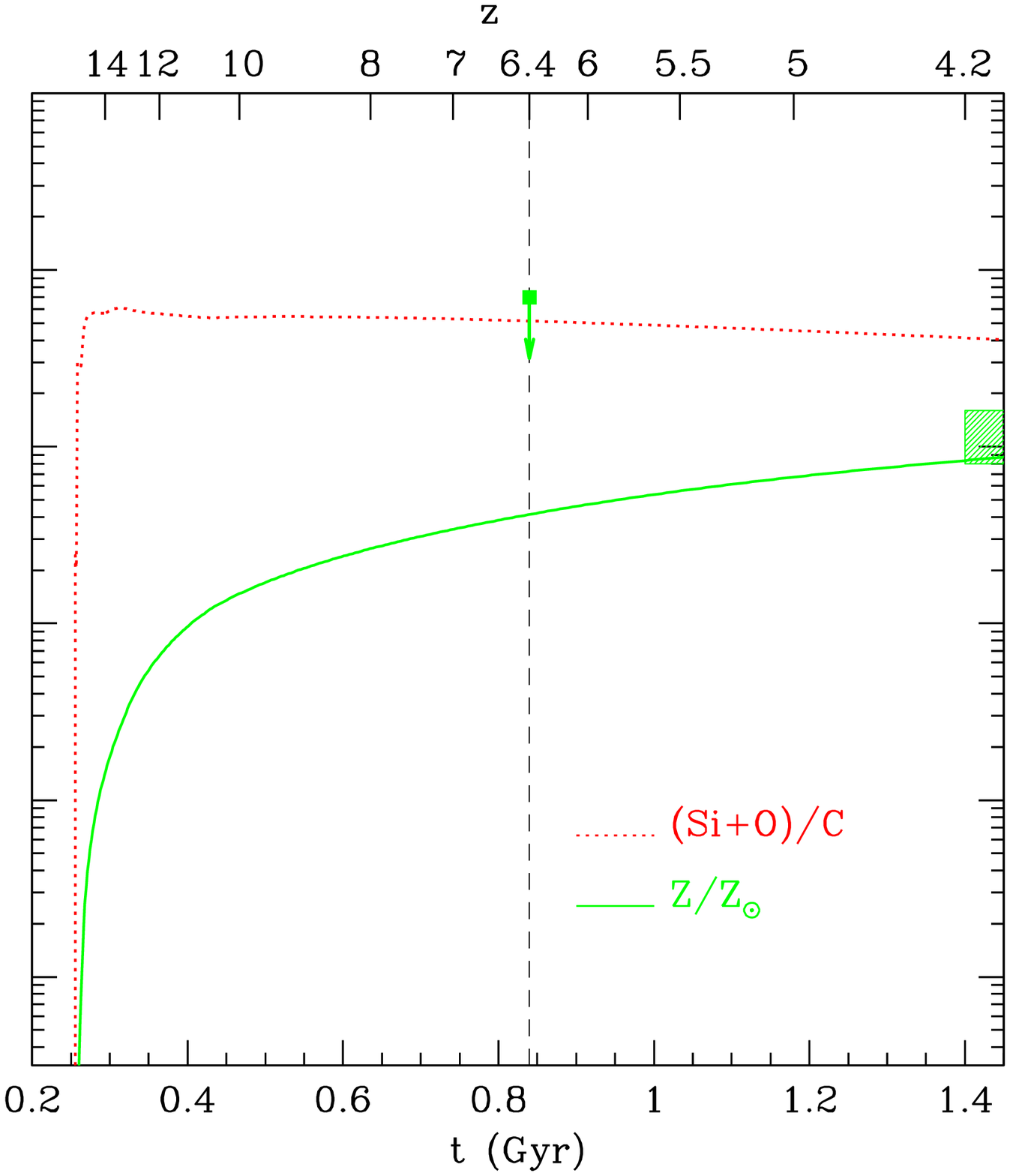}
\caption{Chemical evolution of the high redshift quasar SDSS J1148+5251 corresponding to the star formation histories shown in Fig.~\ref{fig:Lisfr}
(left panel, $\rmn{SFR}$ simulated by Li et al. (2007); right panel, constant $\rmn{SFR}$ with the average value of $1000 M_{\odot}$yr$^{-1}$). 
The solid line represent the evolution of the metallicity in solar units. The upper limit is
derived from the metallicity measured in the BLR (Juarez et al. 2009) and the shaded area
indicates the range of values derived from NLRs of a sample of lower-$z$ radio-galaxies 
(Matsuoka et al. 2009). The dotted line represent the evolution of the (Si+O)/C  elemental ratio. 
The vertical dashed line indicates the Hubble time corresponding to $z=6.4$. 
}
\label{fig:Liqsomet}
\end{figure*}

In Fig. \ref{fig:Liqsomet} we show the predicted evolution of the total metallicity 
in solar units (solid line) for the two star formation models shown in Fig. \ref{fig:Liqso}. 
The shaded area indicates the metallicity level inferred from observations of the narrow 
line regions (NLR) of a sample of high-$z$ radio-galaxies, which shows no evolution over 
the redshift range  $1 \leq z \leq 4$ (Matsuoka et al. 2009). Juarez et al. (2008)
determined the metallicity of the BLR of J1148+5251. Since the BLR is a small nuclear region, 
of less than a few pc, contaning a total mass of $\sim 10^4 M_{\odot}$, we consider
this measurement to be an upper limit to the metallicity of the host galaxy.
Conversely, the NLR has a size which is roughly comparable to the size of the host galaxy 
and represents a good tracer of the chemical properties on galactic scales (Matsuoka
et al. 2009 and references therein). 

Adopting the star formation history predicted by the simulation of Li et al. (2007, 
left panel), the ISM reaches the solar metallicity already at $z \sim 8$, and remains 
constant thereafter. A delayed metal-enrichment is found if a constant star formation rate 
is assumed (right panel) and the predicted metallicity at $z = 6.4$ is $\sim 0.4 Z_\odot$.  
Both models are consistent with the upper limit to the total metallicity derived from 
observational data for J1148+5251, and are in good agreement with
the data inferred from NLRs observations at $z \sim 4$. 

Finally, we would like to comment on the so-called "Carbon-problem" pointed out by 
Juarez et al. (2009). These authors argue that the lack of evolution of the observed 
(SiIV+OIV)/CIV ratio between $4 \leq z \leq 6.4$ suggests that the abundance of carbon 
relative to silicon and oxygen also does not evolve significantly. If carbon is mostly 
produced by AGB stars, the bulk of its production must be delayed with respect to that 
of $\alpha$-elements and we should expect to see an evolution at $z > 6$. However,   
the dotted lines in Fig.\ref{fig:Liqsomet} show that both star formation models 
predict a (Si+O)/C ratio of $\sim 6$ at $z=6.4$, with only a mild evolution at lower 
redshifts. The evolution is more pronounced in the left panel (a factor 2 over the 
redshift range $4.2 < z < 6.42$) because at $z \leq 6.4$ we expect to see the 
contribution to carbon (and carbon dust) enrichment of intermediate mass stars formed 
in the last major burst of star formation. As already discussed above, the contribution 
of intermediate mass stars is much less important when a constant star formation rate 
is assumed (right panel). The resulting mild evolution indicates that carbon production 
by massive short-lived stars, progenitors of SNe, is not negligible relative to that of 
longer-lived intermediate mass stars, at least according to the stellar metal yields 
adopted in the present paper.   

\section{Discussion and Conclusions}

In this work we show that the role of AGB stars in cosmic dust evolution at high redshift might have been over-looked. We develop  
a chemical evolution model which follows the evolution of metals and dust on the stellar characteristic lifetimes, taking into account
dust destruction mechanisms. We use a grid of stellar dust yields as a function of the initial mass and metallicity over
the range $1 - 40 M_\odot$ and $0 - 1 Z_\odot$. We adopt the stellar dust yields by Zhukovska et al. (2008) for AGB stars with stellar 
masses in the range $(1-7) M_{\odot}$ and the model of Bianchi \& Schneider (2007) to account for dust production and destruction 
in the reverse shock of SNe, with progenitor stellar masses in the range $12 - 40 M_{\odot}$. 

Our results can be summarized as follows:
\begin{itemize}
\item A comparison of dust yields expected from AGB stars and SNe shows that carbon grains are the dominant dust component.
      Silicates and other dust components can only be produced by AGB stars with solar metallicities and by SNe when $Z \geq 10^{-2} Z_\odot$
      but are always sub-dominant with respect to carbon grains. This is at odd with what has been generally adopted in chemical evolution
      models which, assuming a dust condensation efficiency of 1 for all refractory elements present in the ejecta, predict that SNe dominate
      silicate dust production and that AGB stars lead to a delayed injection of carbon dust (Dwek 1998; Dwek 2005; Galliano, Dwek \& Chanial 2008).
\item At early times, dust production is dominated by SNe as a consequence of the shorter lifetimes of their progenitor stars.
      AGB stars start to produce dust after about 30~Myr (i.e. when a $8 M_{\odot}$ star evolves off the main sequence 
      to the AGB phase). For our reference model, where we adopt a Larson IMF with $m_{ch} = 0.35 M_{\odot}$ and stellar lifetimes from 
      Raiteri et al. (1996), we find that the characteristic timescale at which AGB stars dominate dust production ranges between 150 and  500~Myr, 
      depending both on the assumed star formation history and on the initial stellar metallicity. Hence, we conclude that these stellar dust sources 
      must be taken into account at early cosmic epochs. 
\item This conclusion is only moderately dependent on the adopted stellar lifetimes, but it is significantly affected by variations of the IMF: for a
      $m_{ch} = 5 M_\odot$, dust from AGB starts to dominate only on timescales larger than 1 Gyr, and SNe are found to dominate dust evolution when
      $m_{ch} \geq 10 M_\odot$. Thus, variations of the stellar characteristic mass over cosmic history due to the smaller gas metallicity and/or to the larger
      cosmic microwave background temperature (see, e.g., Omukai et al. 2005) might significantly affect the origin and properties of dust in the high redshift Universe.
\item We apply the chemical evolution model with dust to the host galaxy of the most distant QSO at $z = 6.4$, SDSS J1148+5251. Given the current uncertainties on the 
      star formation history of the host galaxy, we have considered two models: (i) the star formation history obtained in a numerical simulation by Li et al. (2007) 
      which predicts that a large stellar bulge is already formed at $z = 6.4$, in agreement with the local SMBH-stellar mass relation, through a series of strong bursts 
      triggered by galaxy mergers in the hierarchical evolution of the system and (ii) a constant star formation rate of $1000 M_\odot$/yr, as suggested by the observations 
      if most of the FIR luminosity is due to young stars. In the latter case, the stellar mass at $z =6.4$ is compatible with estimates of the dynamical mass derived by 
      observations. The total mass of dust predicted at $z = 6.4$ by the first model is $2 \times 10^8 M_\odot$, within the range of values inferred by 
      observations, with a substantial contribution ($\sim 80$\%) of AGB-dust. When a constant star formation rate is adopted, the contribution of AGB-dust decreases 
      to $\sim 50$\% but the total mass of dust formed is a factor 2 smaller. Both models predict a rapid enrichment of the ISM with metals and a relatively mild evolution 
      of the carbon abundance, in agreement with observational constraints. 
\end{itemize}

The evolution of dust in the high-redshift Universe is still very uncertain. Some of the chemical evolution models with dust developed so far
have been aimed at reproducing the elemental abundances of the gas and dust phases of the ISM of our galaxy (Dwek 1998; Falceta-Goncalves 2008; 
Zhukovska et al. 2008) and of local ellipticals and dwarf irregulars (Calura, Pipino \& Matteucci 2008). 
In early galaxies, a crucial unknown is the relative contribution of SNe and AGB stars. 
It is generally found that there is no difficulty in producing highly dusty galaxies at $z \geq 5$
if SNe are an important contributor to interstellar dust. However, the models developed so far assume that all refractory elements
present in SN ejecta form dust grains with a condensation efficiency of 1 (Dwek et al. 2007) or neglect dust destruction mechanisms
(Morgan \& Edmunds 2003). 

The evolution of metals and dust in the host galaxy of the most distant quasar at $z =6.4$ SDSS J1148+5251 has been also investigated  
by Dwek et al. (2007) which show that an average SN must condense a dust mass of $1 M_{\odot}$ to reproduce the dust-to-gas ratio inferred
from the observations (0.0067), when a gas mass of $3 \times 10^{10} M_{\odot}$ and a dust mass of $2 \times 10^8 M_{\odot}$ are assumed.
Since the observed dust yield is much lower (the largest observed dust yield is $\sim 0.054 M_{\odot}$, Rho et al. 2008), the authors 
suggest that MC-grown dust or dust formed around the AGN (i.e. in outflows from the broad line regions, Elvis et al. 2002) must be taken into account.
Our results suggest the alternative possibility that AGB stars might significantly contribute to dust production even at redshifts $z \geq 6$. 
For a comparison, we can infer from figure 8 in Dwek et al. (2007) that, in order to reproduce the 
dust-to-gas ratios predicted by our SFR models for SDSS J1148+5251 ($(1 - 6.7) \times 10^{-4}$ at $z = 6.4$), a dust yield per SN of about 
$0.1 M_{\odot}$ would be required. This is larger than our adopted SN yields, the difference being compensated by the additional contribution to
the dust budget provided by AGB stars.
 
More recently, Li et al. (2008) have applied a three-dimensional Monte Carlo radiative transfer code to the hydrodynamic simulations 
of Li et al. (2007) to model the dust properties of J1148+5251 and find that a supernova-origin dust model may be able to explain the 
observed dust properties. Yet, dust destruction in the ISM is not explicitely taken into account. 
Moreover, these models make the a-priori assumption that at redshift above 6, when
the age of the Universe is less than 1 Gyr, the major alternative source of interstellar dust, namely AGB stars, have not yet evolved
to their dust-production stage.      

Our results show that stellar sources can account for the huge dust masses inferred from observations of
distant QSOs. Although the star formation history of the QSO host galaxies is still largely unknown, our analysis indicate that 
the contribution of AGB stars to dust production cannot be neglected, even at the most extreme redshifts currently accessible to
observations. Yet, since AGB winds are not efficient to disperse dust through the galaxy, SNe are still required to efficiently
re-distribute dust on galactic scales (Jones 2005). 

The nature and properties of the grains responsible for dust extinction in high-redshift QSOs and GRBs host
galaxies (Maiolino et al. 2004; Stratta et al. 2007; Gallerani et al. in prep) are likely to depend more on the specific host galaxy 
star formation history than on redshift. Future observations aimed at constraining the properties of dust in high redshift QSOs 
might be a powerful probe of star formation and QSO feedback regulating the BH-galaxy coevolution.

\section*{Acknowledgments}
We thank the anonymous referee for the careful reading and the insightful comments which helped us to improve the clarity
of the paper. We are grateful to A. Ferrara, S. Gallerani, H. Gomez, A. Jones, R. Maiolino, D. Maoz, A. Marconi and S. Salvadori for 
useful suggestions. R. Schneider acknowledges the support and hospitality of the Dark Cosmology Center. 
The Dark Cosmology Centre is funded by the Danish National Research Foundation.

\label{lastpage}

\end{document}